\documentclass[english,11pt]{article}
          \usepackage{babel}
          \usepackage[T1]{fontenc}
          \usepackage[latin2]{inputenc}
          \usepackage{pslatex}
          \usepackage{epsf}
\def\refitem#1\par{\parindent=0truemm\parskip=0pt\hangindent=2truecm
     \frenchspacing#1\par}

\def \AA #1 #2 {{\em Astron.Astrophys.,\/} {\bf #1}, #2}
\def \ApJ #1 #2 {{\em Astrophys.J.,\/} {\bf #1}, #2}
\def \Acta #1 #2 {{\em Acta Astr.,\/} {\bf #1}, #2}
\def \AJ #1 #2 {{\em Astron.J.,\/} {\bf #1}, #2}
\def \PASP #1 #2 {{\em Publ.Astr.Soc.Pacific,\/} {\bf #1}, #2}
\def \MNRAS #1 #2 {{\em M.N.R.A.S.,\/} {\bf #1}, #2}

\begin{document}

\title{\bf Are Disks in Dwarf Novae during their Superoutbursts  
\\Really Eccentric ?}

\author{ J. Smak } 
\date{ Nicolaus Copernicus Astronomical Center,~ Polish Academy of Sciences \\
ul.~Bartycka~18,~ 00-716~Warsaw, Poland \\
{ e-mail: jis@camk.edu.pl} }

\maketitle

\begin{abstract}

The evidence presented earlier by several authors for the substantial 
disk eccentricity in dwarf novae during their superoutbursts is shown 
to result either from errors, or from arbitrary, incorrect assumptions. 

(1) The evidence for Z Cha and WZ Sge (Vogt 1981), based on 
radial velocities measured from absorption components, was an artifact, resulting from miscalculated beat phases.   

(2) The evidence for OY Car (Krzemi{\'n}ski and Vogt 1985) and 
IY UMa (Patterson et al. 2000), based on the observed dependence 
of eclipse parameters on the beat phase, involved an implicit 
assumption that the observed eclipses are {\it pure disk} eclipses, 
which is not true. In particular, the observed variations of eclipse 
parameters are likely due to the contributions from the hot spot 
and from the superhump source, which depend strongly on the beat phase. 

(3) The evidence for OY Car (Hessman et al 1992) and WZ Sge 
(Patterson et al. 2002), resulting from the analysis of hot spot 
eclipses, was based on the assumption that the spot distances 
are identical with the radius of the disk, which is not always 
correct. In particular, in the case of eclipses of "peculiar" spots   
(involving the stream overflow), observed at beat phases away from 
$\phi_b \sim 0.5$, the resulting spot distances are smaller that 
the radius of the disk. 

New determination of disk eccentricity in Z Cha, using Vogt's 
radial velocities measured from emission components, 
gives $e=0.05 \pm 0.05$. 

\end{abstract}

\noindent
{\bf Key words:}
{\it accretion, accretion disks -- binaries: cataclysmic variables,
stars: dwarf novae, stars: individual: OY Car, Z Cha, WZ Sge, IY UMa }

\section {Introduction }

The concept of an eccentric, "precessing" 
\footnote 
{ The term "precession", used commonly by many authors, 
is obviously incorrect. We will use it -- for simplicity -- only 
in the Introduction but will revert later to the correct term: 
the "apsidal motion". } 
disk was a crucial ingredient of the tidal-thermal instability model 
of superoutbursts proposed by Osaki (1996, 2005 and references therein). 
Results of recent analysis of superoutbursts of Z Cha (Smak 2007, 2008a)
definitely show, however, that superoutbursts are due to a major
enhancement in the mass transfer rate.
It is worth to recall that an alternative model for superoutbursts, 
involving such enhanced mass transfer resulting from variable irradiation 
of the secondary component, was considered earlier by Osaki himself 
(Osaki 1985; see also Mineshige 1988). 
Further work in this direction appears now highly desirable.

Eccentric, "precessing" disk continues to be an essential ingredient
of the commonly accepted model for superhumps (see also Osaki 1996, 2005
and references therein).

Observational evidence for eccentric, "precessing" disks in dwarf novae
during their superoutbursts was presented during the last three decades
by numerous authors. Under those circumstances the question posed in the
title of the present paper can be considered rather provocative.
Even more provocative, however, will be our negative answer to that question.

The organization of the present paper will be as follows. 
In Section 2 we will clarify problems of terminology. 
The main part of the paper will then be devoted to a detailed critical 
review of the observational evidence for eccentric, "precessing" disks 
resulting from the analysis of radial velocities (Section 3), 
of the observed (disk?) eclipses (Section 4), and of the spot of eclipses 
(Section 5). Results will be discussed in Section 6.

\section {Terminology and Definitions }

Three periodicities are observed during superoutbursts:
the orbital period $P_{orb}$, the superhump period $P_{sh}$,
and the corresponding beat period $P_b$

\begin{equation}
{1\over {P_b}}~=~{1\over {P_{orb}}}~-~{1\over {P_{sh}}}~.
\end{equation}

Accordingly, three sets of phases are to be considered:
the orbital phase $\phi_{orb}$ (or simply $\phi$), 
the superhump phase $\phi_{sh}$, and the beat phase

\begin{equation}
\phi_b~=~\phi_{orb}~-~\phi_{sh}~~=~\phi_{orb}({\rm SH})~
	 =~-~\phi_{sh}({\rm ecl})~.
\end{equation}

It is commonly assumed that the beat period $P_b$ represents the 
apsidal motion of an eccentric disk (or an outer eccentric ring),
usually referred to -- incorrectly -- as "precession".
Consequently, the beat period is called the "precession" period while
the beat phase -- the "precession" phase.

The orientation of an eccentric disk/ring can be defined by the position 
angle $\theta_{\circ}$ of the periastron with respect to the line of sight.
This position angle is directly related to the beat phase $\phi_b$.
In general, they could differ by a constant phase shift.
In practice, however, using various kinds of observational evidence,
nearly all authors (e.g. Vogt 1981, Hessman et al. 1992) 
assume or conclude that the beat phase $\phi_b=0$ corresponds to 
the situation when the periastron is facing the observer, i.e. that

\begin{equation}
\theta_{\circ}~\equiv ~\phi_b~.
\end{equation}

\section { Evidence from Radial Velocities }

\subsection { Radial Velocities from an Eccentric Ring }

Using textbook formulae for the two body problem we find that the
radial velocity of a point on an elliptical ring is given by

\begin{equation}
V_r~=~V_r({\rm orb})~+~V_d~\sin i~
     \left [~\sin(\theta_{\circ}+\theta)~+~e~\sin \theta_{\circ})~\right ]~,
\end{equation}

\noindent
where $V_r({\rm orb})$ is describing the orbital motion,
$\theta$ is the position angle of the point considered with respect to
the periastron, $\theta_{\circ}$ -- the position angle of the periastron with
respect to the line of sight (both counted counterclockwise), and

\begin{equation}
V_d~=~\left [ {GM_1}\over{A_d~(1-e^2)} \right ]^{1/2}~,
\end{equation}

\noindent
is an equivalent of the standard expression for the rotational
velocity of the disk/ring. Here $A_d$ is the major semi-axis of the ring
and $e$ -- its eccentricity.

For the radial velocity of an element of the ring seen in projection
on the white dwarf and the central parts of the disk
(to be discussed in Sections 3.2 and 3.3) we obtain

\begin{equation}
V_r~=~V_r({\rm orb})~+~V_d~\sin i~e~\sin \theta_{\circ}~.
\end{equation}

For the maximum and minimum radial velocity, corresponding to the red
and blue peaks of the emission line (to be used in Section 3.4) we get

\begin{equation}
V_r~=~V_r({\rm orb})~+~V_d~\sin i~(~\pm~1~+~e~\sin \theta_{\circ})~.
\end{equation}

\subsection { Z Cha -- absorption lines }

The first evidence for an eccentric disk in a dwarf nova during
the superoutburst came from spectroscopic observations of Z Cha made by
Vogt (1981) during its 1978 March/April superoutburst. Vogt measured
radial velocities from the central absorption components of the double
emission lines and found (see his Fig.7) that those measured
on 1978 March 28/29 were very high (his $\gamma_{28}=+267\pm 13$ km/s)
while those measured on 1978 March 27/28 (i.e. on the previous night)
-- very low ($\gamma_{27}=-197\pm 37$ km/s).

To intepret those results Vogt proposed the following model
(see Fig.9 of his paper): The central absorption components are produced
in the outer eccentric ring, specifically --
in the region which is seen in projection on the central parts of the disk.
The difference between $\gamma_{27}$ and $\gamma_{28}$ results from
the apsidal motion of the ring: the value of the beat period
$P_b\approx 2.1$d implies that the orientation of the eccentric disk
on two consecutive nights differs by about 180$^{\circ}$.

This interpretation, however, can be easily shown to be incorrect.
To do so we calculate the beat phases corresponding to the times
of Vogt's spectroscopic observations. For the night of March 27/28 
we obtain $\phi_b=0.39$, while for the night of March 28/29 we get 
$\phi_b=0.83-0.90$.
From Eq.6 it immediately follows that radial velocities measured
on March 27/28 should be systematically {\it positive}, while those
on March 28/29 -- systematically {\it negative}.
This, however, is just opposite to the observed effect. 
In other words, the orientations of the ring during those two 
nights were -- roughly -- opposite to the two orientations 
shown in Vogt's Fig.9. 

To complete our discussion let us return to the problem of the origin
of central absorption components. Vogt assumed -- arbitrarily --
that they are formed in the outer eccentric ring. 
In reality, however, the double emission lines, including their central
absorption components, are produced in the atmosphere of the entire disk.
Furthermore, in the particular case of Z Cha, an additional, variable
contribution due to the absorption in the overflowing parts of the stream
(Smak 2007) is likely to be present. Such a possibility is suggested
by the fact that, according to Vogt, the absorption components are 
strongest near the orbital phase $\phi_{orb}=0.9$, i.e. when 
the overflowing stream is directly projected against the central 
parts of the disk.

\subsection { WZ Sge -- absorption lines }

To strengthen his interpretation of Z Cha Vogt presented additional evidence
based on absorption-line $\gamma$-velocities of WZ Sge measured by several
authors during its 1978 superoutburst. 
His Fig.8 did indeed show sinusoidal variations of the $\gamma$-velocities with the beat phase. 
However, to calculate $\phi_b$ Vogt adopted an incorrect value of
the beat period: $P_b=6.176$d. 
The corrected version of his $\gamma~vs.\phi_b$ diagram, with $\phi_b$
calculated with $P_b=7.16$d (Patterson et al. 1981), shows only large scatter with {\it no} obvious dependence of $\gamma$-velocities 
on $\phi_b$.

\subsection { Z Cha -- emission lines }

The eccentricity of the disk can be determined, using Eq.7, from Vogt's 
radial velocities measured from the peaks of the double emission lines. 
For this purpose we use radial velocities E$^+$ and E$^-$ listed
in his Table 6 (excluding only four uncertain values of E$^+$ with
$\phi_{orb}$ between 0.10 and 0.40 which were near-blends with the
absoprtion component). First, we correct them for the orbital motion 

\begin{equation}
V_r({\rm orb})~=~\gamma~-~K_1~\sin~(\phi~-~\phi_{\circ})~,
\end{equation}

\noindent
using two, rather different sets of parameters determined from 
observations at quiescence, namely: 
(1) $K_1=87$ km/s, $\phi_{\circ}=-0.020$, $\gamma=0$ km/s (Vogt 1981), 
and (2) $K_1=192$ km/s, $\phi_{\circ}=0.098$, $\gamma=2$ km/s 
(Marsch et al. 1987).

Results, obtained with those two sets of parameters, are:
(1) $V_d\sin i=650\pm21$ km/s, $e=0.05\pm0.05$, $r_d=0.54\pm 0.04$,
and (2) $V_d\sin i=649\pm21$ km/s, $e=0.02\pm0.05$, $r_d=0.53\pm 0.03$.
As can be seen the eccentricity, if any, is close to zero.
Worth noting is that in both cases the resulting disk radius turns out, 
as expected, to be close to the mean radius of the Roche lobe 
$r_{Roche}=0.52$.

\section { Evidence from the Observed (Disk?) Eclipses }

\subsection { General comments }

The evidence discussed below for OY Car (Section 4.2) and IY UMa (Section 4.3)
was based on the dependence of various eclipse parameters on the beat phase, 
interpreted in terms of the apsidal motion of an eccentric disk. 
Such an interpretation, however, assumes implicitly that the observed eclipses 
are {\it pure disk} eclipses. This is obviously not true. 

We now have evidence, based on the detailed analysis of eclipses in Z Cha
(Smak 2007), showing that, depending on the beat phase, 
they represent a combination of eclipses of several different sources: 
(1) the disk, (2) the "standard" spot (around $\phi_b \sim 0.5$), or
(3) the "peculiar" spot, involving stream overflow 
(around $\phi_b \sim 0.25$ and 0.75), and 
(4) the superhump source (around $\phi_b \sim 0.0$). 
Their relative contributions depend strongly on the beat phase.
Consequently, the resulting dependence of the observed eclipse parameters
on $\phi_b$ is quite complicated.
Under those circumstances the only practical solution is to decompose
the observed light curves into their separate components (as it was done
in the case of Z Cha; Smak 2007). 
It can be hoped that the existing light curves of OY Car and IY UMa
will eventually be analyzed in such a way. 
For the time being we can only mention that the shapes of the {\it pure disk}  
eclipse light curves obtained for Z Cha (Smak 2007) do not show 
any dependence on the beat phase.

\subsection { OY Car }

Krzemi{\'n}ski and Vogt (1985) collected light curves of OY Car during
its January 1980 superoutburst and determined several parameters describing
the shape of the observed eclipses:
(1) deviations of the eclipse amplitude $\Delta A$ from the
amplitude $A$ vs. epoch $E$ relation,
(2) deviations of the eclipse width $\Delta W$ from the width $W$ vs.
epoch $E$ relation,
(3) the eclipse asymmetry $\Delta T$, measured at half-depth, and
(4) the values of ($O-C$).
They found that all those parameters show -- roughly -- sinusoidal variations
with the beat phase (see their Figs.7-11) and interpreted them as being
due to the apsidal motion of an eccentric disk.
In what follows we will show that this interpretation was incorrect.

First of all, as noted by Krzemi{\'n}ski and Vogt themselves, variations
predicted by their eccentric disk model (see their Fig.12) for
$\Delta W$ and $\Delta T$ do not agree with their observed variations.
Specifically, their model predicts {\it double sinusoidal} variations for
$\Delta W$, while in the case of $\Delta T$ the model predicted variations
are shifted in $\phi_b$ by $\pm 0.25$.

To interpret the observed variations of $\Delta A$ 
Krzemi{\'n}ski and Vogt assumed that the outer parts of the disk,
on the other side of the white dwarf, remain uneclipsed.
Using system parameters of OY Car (Wood et al. 1989) 
we find, however, that this assumption would require the disk radius
to be larger than $r=0.78$, which is already larger than
the upper limit to the disk radius $r_{max}=0.72$, obtained from the condition
that it must be smaller than the distance from L$_1$ to the white dwarf.
This implies that the far side of disk is fully eclipsed.
The geometry of eclipses of the hypothetical eccentric disk in OY Car
is shown schematically in Fig.1.
It can immediately be seen that, regardless of the surface brightness
distribution, the eclipse amplitude should show {\it double sinusoidal}
variations with $\phi_b$, which is inconsistent with observations.

\begin{figure}[h]
\bigskip
\epsfysize=1.5cm 
\hspace{1.5cm}\epsfbox{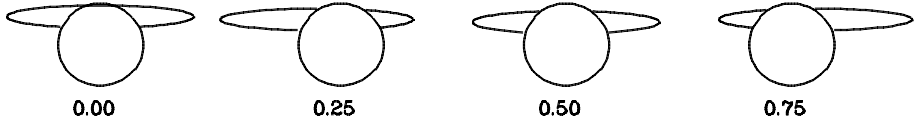} 
\caption[h]{Geometry of eclipses of a hypothetical eccentric disk in OY Car at
$\phi_{orb}=0$ and at beat phases $\phi_b=0.0$, 0.25, 0.5, and 0.75.} 
\bigskip
\end{figure}

The observed variations of $\Delta A$ with $\phi_b$ can, in fact,
be qualitatively explained by the variable contribution from the hot spot.
In particular, around $\phi_b \sim 0.5$, when the "standard" hot spot is
present, its eclipses make the full eclipse amplitude larger.

\subsection { IY UMa }

Patterson et al. (2000) presented results of their extensive photometry
of IY UMa covering the declining part of its 2000 January superoutburst.
From the analysis of the observed light curves they found that
all eclipse parameters depend strongly on the beat phase.
In particular, around $\phi_b \sim 0.5$ the eclipse amplitude is largest
and all characteristic phases are systematically more positive
(see their Figs.7 and 9). Patterson et al. interpreted those effects
as being to due the apsidal motion of a strongly eccentric ($e=0.29$) disk.

There are several arguments against such an interpretation.
First of all, we can recall our general comments from Section 4.1.
Secondly, we may note that the observed dependence of eclipse parameters
on $\phi_b$ (similar to the case of OY Car; Section 4.2) is most likely
due to the variable contribution from the hot spot which --
on account of the orbital inclination ($i=86.8$) being much higher than
in the case of OY Car ($i=83.3$) -- is expected to be relatively larger.
Around $\phi_b \sim 0.5$, when the "standard" hot spot is present,
the contribution from its eclipses makes the observed eclipses deeper
and shifted toward later phases. 
In partcular, the large values of $\phi_4$ observed around $\phi_b \sim 0.5$ 
(Fig.9 in Patterson et al. 2000) correspond probably to the egress 
of the "standard" hot spot.

\section { Evidence from Spot Eclipses }

\subsection { Historical comments }

The presence of the hot spot in a dwarf nova during its superoutburst
was detected for the first time by Schoembs (1986). His light curves
of OY Car obtained during the final decline from its 1980 November/December
superoutburst showed clearly the characteristic eclipses of the hot spot.
Six years later Hessman et al. (1992) published results of their
photometry of OY Car, made also during the final decline from another of its superoutbursts, their light curves showing also characteristic spot eclipses.
Unfortunately, the significance of these discoveries was not fully appreciated
at that time, the presence of those eclipses being interpreted in terms
of the {\it reappearance} of the quiescent hot spot.

The hot spot eclipses during superoutburst maximum were first detected
by Patterson et al. (2002) in WZ Sge during its 2001 superoutburst.
Regretfully, this interpretation was challenged by Osaki and Meyer (2003)
who argued that those eclipses were not due to the hot spot 
but rather due to {\it "the superhump light source itself"}. 

Situation was definitely clarified only very recently by the results of
our analysis of eclipse light curves of Z Cha and OY Car (Smak 2007,2008b)
observed during maxima of their superoutbursts.
They were decomposed into their disk and spot components, the resulting hot
spot light curves showing not only the characteristic shape of the eclipse
but also the characteristic maximum around the orbital phase 
$\phi \sim 0.8-0.9$.

\subsection { How Reliable are Disk Radii derived from Spot Eclipses? }

According to the commonly accepted "standard" model, the hot spot is 
produced by the collision of the stream with the outer parts of the disk.
In such a case, the distance $r$ of the spot from the white dwarf,
to be referred to as "spot distance", determined from spot eclipses,
is identical with the radius of the disk $r_d$.
It is worth to emphasize that in the case of an eccentric disk this $r_d$
is the {\it local} radius of the disk at a specific position angle
corresponding to the point of collision. 
Furthermore, when such an eccentric disk rotates due to the apsidal 
motion, the dependence of $r_d$ on the position angle produces 
its periodic variations with the beat phase. 

In the analysis of eclipses of the "standard" hot spot we use the phases
of mid-ingress $\phi_i$ and mid-egress $\phi_e$ which are related to the mass
ratio $q$ (defining the shape of the stream trajectory),
the orbital inclination, the spot distance $r$,
and its ellongation parameter $\Delta s$ (cf. Smak 1996,2007).
When the mass ratio and inclination are known, the two phases $\phi_i$
and $\phi_e$ can then be used to obtain two indpendent determinations
of the spot distance:

\begin{equation}
r_i(\phi_i)~=~f(\phi_i,\Delta s)~~~{\rm and}~~~
		r_e(\phi_e)~=~f(\phi_e,\Delta s)~.
\end{equation}

\noindent
It is obvious that in the case of a "standard" hot spot they must be
identical

\begin{equation}
r_i(\phi_i)~=~r_e(\phi_e)~.
\end{equation}

The observed luminosity of the hot spot depends on the "impact parameter"
$\Delta V^2$ (i.e. on the square of the relative velocity of collision
between the stream and the disk).
The value of this impact parameter increases with decreasing radial
distance $r$. Therefore in the case of variable disk radius the luminosity
$\ell_s$ of the "standard" spot should vary as 

\begin{equation}
{{d\ell_s}\over{dr}}~<~0~.
\end{equation}

The last two relations provide two crucial tests for the applicability of
the concept of a "standard" hot spot and -- in particular -- for assuming
that the radial distance of the spot ($r_i$ or $r_e$) is identical with the
radius of the disk $r_d$.

\begin{figure}[h]
\bigskip
\epsfysize=9.0cm 
\hspace{4.5cm}\epsfbox{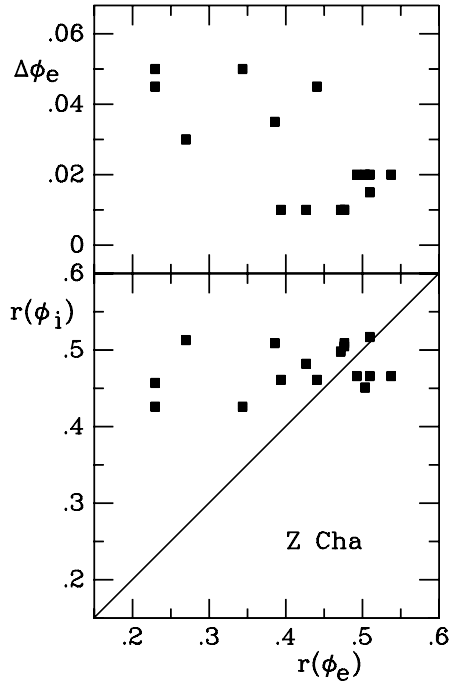} 
\caption[h]{Results of the analysis of spot eclipses in Z Cha during its superoutbursts.
{\it Bottom}: Spot distances $r(\phi_i)$ determined from ingress 
are compared with spot distances $r(\phi_e)$ determined from egress.
{\it Top}: Durations of egress $\Delta \phi_e$ are plotted 
against spot distances $r(\phi_e)$.}
\bigskip
\end{figure}

Recent analysis of hot spot eclipses in Z Cha during its superoutbursts
(Smak 2007) provides good illustration of different situations to be
encountered in this area.
The "standard" hot spot eclipses are observed only at beat phases 
$0.40<\phi_b<0.60$. In particular, the spot distances $r_i$ and $r_e$ 
determined from those phases are practically identical.
On the other hand, the spot eclipses observed at intermediate beat
phases (around $\phi_b\sim 0.35$ and around $\phi_b\sim 0.75$) 
are rather peculiar. 
First of all, the resulting spot distances are generally smaller,
with $r_i$ being systematically larger than $r_e$. 
This could be seen already from Fig.6 of Smak (2007) and is shown here 
in the lower part of Fig.2. 
Secondly, the luminosities of the spot observed at intermediate beat
phases are systematically lower (Smak 2007, Fig.5).

Those peculiarities were interpreted (Smak 2007) as being due to 
a substantial stream overflow. Supporting this interpretation is 
a correlation between the duration of egress $\Delta \phi_e$ 
and the spot distance $r_e$, mentioned already in Section 5.3 of 
that paper, and shown here in the upper part of Fig.2. 

The lesson learned from Z Cha can be summarized 
in the form of the following warning: The spot distances $r_i > r_e$,
obtained from the analysis of peculiar spot eclipses, observed
at intermediate beat phases, {\it do not provide any reliable 
information about the radius of the disk}. 
This warning becomes even stronger when the observed spot 
luminosities imply $d\ell_s/dr > 0$.

\subsection { OY Car }

As already mentioned above, Schoembs (1986) and Hessman et al. (1992)
presented extensive photometric coverage of the advanced stages of
the 1980 November/December and 1987 February superoutbursts of OY Car.
During the final decline their light curves showed characteristic
eclipses of the hot spot.
Hessman et al. (1992) analyzed those light curves in order to determine
the location of the hot spot. Specifically, using the phases
of mid-ingress, they obtained spot distances $r_i$ and found that
they show strong dependence on the beat phase $\phi_b$,
interpreted by them as being due to the "precession"
of a strongly eccentric ($e=0.38$) disk.
This interpretation, however, can easily be challenged.

To begin with, we must note that the variations of the spot luminosity
$\ell_s$ and the spot distance $r_i$ with $\phi_b$, shown in their Fig.8
are strikingly similar to the case of Z Cha (cf. Section 5.2; see also
Figs.5 and 6 in Smak 2007).
This immediately suggests that -- like in the case of Z Cha --
the "standard" hot spot was present in OY Car only around $\phi_b \sim 0.5$
and that spot distances $r_i$ obtained from intermediate beat phases
should not be treated as a reliable measure of the disk radius $r_d$.

\begin{figure}[h]
\bigskip
\epsfysize=7.5cm 
\hspace{5.0cm}\epsfbox{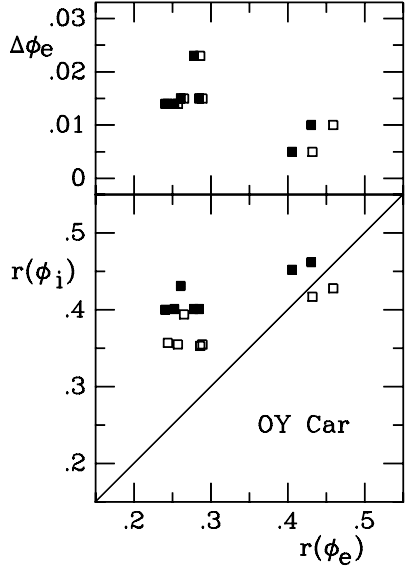} 
\caption[h]{Results of the analysis of spot eclipses in OY Car during its superoutbursts.
{\it Bottom}: Spot distances $r(\phi_i)$ determined from ingress are compared
with spot distances $r(\phi_e)$ determined from egress.
{\it Top}: Durations of egress $\Delta \phi_e$ are plotted against
spot distances $r(\phi_e)$.
Open squares represent results obtained with spot ellongation parameter
$\Delta s=0.02$, while filled squares -- with $\Delta s=0.04$.}
\bigskip
\end{figure}

To clarify the situation definitely we analyze the available light curves 
of seven spot eclipses (eclipses 7-1, 7-3, 9-1, 9-2, 10-1, and 10-2 
from Fig.3b of Schoembs 1986, and the eclipse from Fig.1-bottom 
of Hessman et al. 1992).
For each of them we determine the phases of mid-ingress $\phi_i$ and
mid-egress $\phi_e$ and the duriation of egress $\Delta \phi_e$.
Then, using geometrical system parameters (Wood et al. 1989) 
and adopting two different values of the spot ellongation parameter
$\Delta s=0.02$ and 0.04, we obtain the spot distances $r_i$ and $r_e$.
Results, presented in Fig.3, show that the situation is indeed
nearly identical with that in the case of Z Cha (see Fig.2).
In addition, we have $d\ell_s/dr_i > 0$ (Fig.8 in Hessman et al. 1992).
Taking all this into account we must conclude that
at intermediate beat phases away from $\phi_b \sim 0.5$ we are dealing
with a "peculiar" rather than "standard" spot and that -- consequently --
the spot distances $r_i$ obtained at those phases do not provide
any reliable information about the disk radius $r_d$.

\subsection { WZ Sge }

Patterson et al. (2002) published results of a worldwide
photometric campaign covering the 2001 superoutburst of WZ Sge.
One of their most important results was the detection -- for the first
time -- of hot spot eclipses during the main part of superoutburst maximum.
They analyzed those eclipses in a standard way finding that all hot spot
parameters varied significantly with the "precession" (i.e. beat) phase. 
This was interpreted by them as being due to the apsidal motion 
of a strongly eccentric ($e\geq 0.3$) disk. 
This interpretation, however, can easily be challenged. 

In our analysis we use the values of $(O-C)$ and eclipse 
widths (Fig.21-left of Patterson et al. 2002) to recover 
the phases of mid-ingress ($\phi_i$) and mid-egress ($\phi_e$). 
Then, using system parameters from Steeghs et al. (2007) 
we determine the spot distances $r(\phi_i)$ and $r(\phi_e)$. 
Situation is somewhat complicated here by the fact that the zero-point 
of the orbital phases, defined by mid-eclipse of the spot, does not 
correspond to the true zero-phase. 
The values of the corresponding phase shift, obtained by different authors 
from the analysis of spot eclipses (e.g. Smak 1993) or from radial velocity 
curves (e.g. Spruit and Rutten 1998, Steeghs et al. 2001), range from 
$\Delta \phi =-0.036$ to $\Delta \phi =-0.046$. 
In our analysis we adopt: $\Delta \phi =-0.039~ {\rm and}~ -0.042$. 
For the spot ellongation parameter we also adopt two values: 
$\Delta s = 0.02~ {\rm and}~ 0.04$. 

\begin{figure}[h]
\bigskip
\epsfysize=3.5cm 
\hspace{3.5cm}\epsfbox{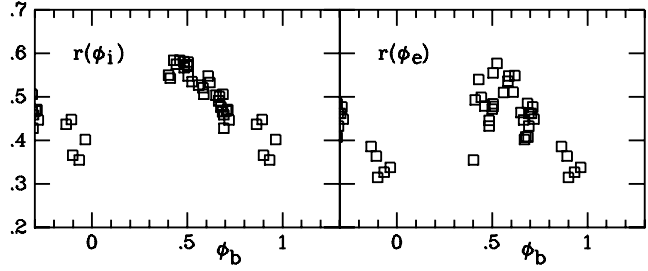} 
\caption[h]{Results of the analysis of spot eclipses in WZ Sge during 
its 2001 superoutburst: 
Spot distances from ingress $r(\phi_i)$ and from egress $r(\phi_e)$,   
determined with $\Delta \phi = -0.039$ and $\Delta s =0.02$, 
are shown as a function of the beat phase. See text for details.}
\bigskip
\end{figure}

\begin{figure}[h!]
\bigskip
\epsfysize=9.0cm 
\hspace{2.5cm}\epsfbox{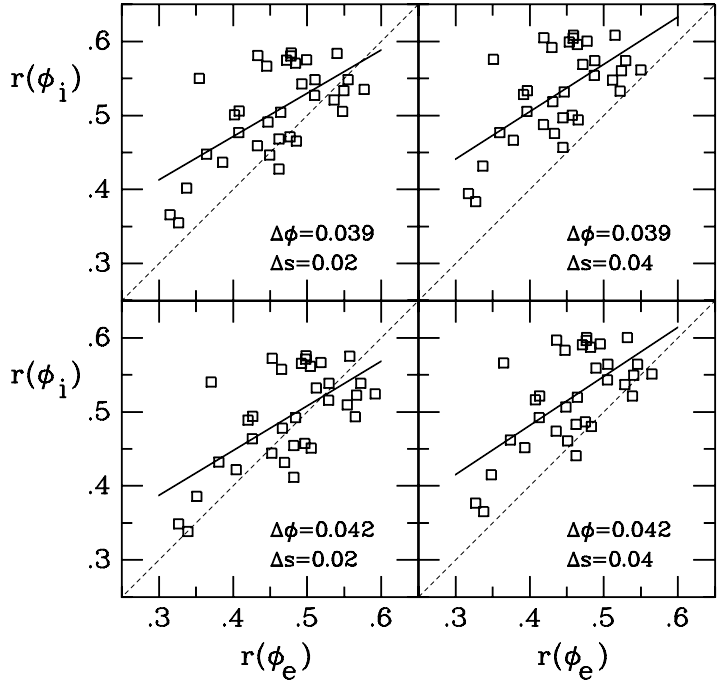} 
\caption[h]{Results of the analysis of spot eclipses in WZ Sge during 
its 2001 superoutburst: 
Comparison of spot distances $r(\phi_i)$ and $r(\phi_e)$ determined   
using four combinations of $\Delta s$ and $\Delta \phi$. 
Solid lines show best fits to the points. }
\bigskip
\end{figure}

Fig.4 shows an example of the resulting spot distances $r(\phi_i)$ and 
$r(\phi_e)$ plotted {\it versus} the beat phase and Fig.5 presents 
a comparison of those two parameters obtained using four combinations of $\Delta s$ and $\Delta \phi$. 
As can be seen, the situation is very similar to the case of Z Cha 
(Fig.2) and OY Car (Fig.3). 
In particular, the condition $r(\phi_i)=r(\phi_e)$ is fulfilled only 
by the largest values of those parameters, corresponding to 
$\phi_b \sim 0.5$, while at other beat phases we have 
$r(\phi_i)>r(\phi_e)$. 
In addition, the spot visibility and its luminosity depend strongly 
on $\phi_b$ (Fig.21-top in Patterson et al. 2002), this behavior 
being strikingly similar to the case of Z Cha. 
Taking all this into account we conclude that away from 
$\phi_b \sim 0.5$ we are dealing with a "peculiar" spot and that -- consequently -- the values of $r(\phi_i)$ and $r(\phi_e)$ obtained 
from those phases are not representative for the radius of 
the disk $r_d$.

\section { Discussion }

Results presented above in Sections 3-5 can be summarized in the form 
of the following conclusions:  

\vskip -10 truemm

\begin {itemize} 

\item The spectroscopic evidence for eccentric disks in Z Cha and WZ Sge, 
presented by Vogt (1981), based on radial velocities measured from 
absorption components, is shown to be an artifact, resulting from 
miscalculated beat phases.   

\item The photometric evidence for eccentric disks in OY Car 
(Krzemi{\'n}ski and Vogt 1985) and IY UMa (Patterson et al. 2000), 
based on the observed dependence of various eclipse parameters 
on the beat phase, involved an implicit assumption that the observed 
eclipses are {\it pure disk} eclipses. This assumption was incorrect. 
In particular, the observed variations of eclipse parameters are 
very likely due to the variable contributions from the 
hot spot and from the superhump source, both of them depending strongly 
on the beat phase. 

\item The photometric evidence for eccentric disks in OY Car 
(Hessman et al 1992) and WZ Sge (Patterson et al. 2002), resulting from  
the analysis of hot spot eclipses, was based on an implicit assumption 
that the spot distances are always identical with the radius of the disk. 
\footnote 
{It may be mentioned that a very similar evidence for substantial 
disk eccentricity was presented by Rolfe et al. (2000, 2001) 
for the dwarf nova IY UMa at quiescence and for the nova-like V348 Pup. 
Our critical comments are likely to apply to those two cases as well. } 
This assumption, however, is correct only in the case of eclipses of 
the "standard" hot spot which are observed around $\phi_b \sim 0.5$. 
At other beat phases we are dealing with "peculiar" spots 
(involving the stream overflow), their eclipses giving spot 
distances $r(\phi_i)>r(\phi_e)$ which are generally smaller 
than the radius of the disk. 

\end {itemize}

To summarize: there is no reliable evidence for eccentric disks 
in dwarf novae during their superoutbursts. 
This conclusion is strengthened by our new determination of disk 
eccentricity in Z Cha, based on Vogt's radial velocities measured 
from emission components, which gave: $e=0.05 \pm 0.05$.

\centerline {\bf References }

\bigskip

\refitem Hessman, F.V., Mantel, K.-H., Barwig, H., Schoembs, R. 1992
	 \AA 263 147.

\refitem Krzemi{\'n}ski, W., Vogt, N. 1985 \AA 144 124.

\refitem Marsh, T.R., Horne, K., Shipman, H.L. 1987 \MNRAS 225 551.

\refitem Mineshige, S. 1988 \ApJ 335 881.

\refitem Osaki, Y. 1985 \AA 144 369.

\refitem Osaki, Y. 1996 \PASP 108 39.

\refitem Osaki, Y. 2005 {\it Proc.Japan Academy, Ser.B}, {\bf 81}, 291.

\refitem Osaki, Y., Meyer, F.  2003 \AA 401 325.

\refitem Patterson, J., McGraw, J.T., Coleman, L., Africano, J.L. 1981
	 \ApJ 248 1067.

\refitem Patterson, J.,Kemp, J., Jensen, L., Vanmunster, T., Skillman, D.R.,
	 Martin, B., Fried, R., Thorstensen, J.R.  2000 \PASP 112 1567.

\refitem Patterson, J. with 27 co-authors  2002 \PASP 114 721.

\refitem Rolfe, D.J., Haswell, C.A., Patterson, J.  2000 \MNRAS 317 759.

\refitem Rolfe, D.J., Haswell, C.A., Patterson, J.  2001 \MNRAS 324 529.

\refitem Schoembs, R. 1986 \AA 158 233.

\refitem Smak, J.  1993 \Acta 43 101.

\refitem Smak, J.  1996 \Acta 46 377.

\refitem Smak, J.  2007 \Acta 57 87.

\refitem Smak, J.  2008a \Acta 58 55.

\refitem Smak, J.  2008b \Acta 58 65.

\refitem Spruit, H.C., Rutten, R.G.M. 1998 \MNRAS 299 768. 

\refitem Steeghs, D., Marsh, T., Knigge, C., Maxted, P.F.L.,  
          Kuulkers, E., Skidmore, W. 2001 \ApJ 562 L145. 

\refitem Steeghs, D., Howell, S.B., Knigge, C., G{\"a}nsicke, B.T., 
          Sion, E.M., Welsh, W.F. 2007 \ApJ 667 442. 

\refitem Vogt, N. 1981 \ApJ 252 653.

\refitem Wood,J., Horne, K., Berriman, G., Wade, R.  1989 \ApJ 341 974.

\end{document}